\documentclass[12pt,preprint,aps,showpacs]{revtex4}
\usepackage{epsfig}

\begin{document}

\title{First evidence for different freeze-out conditions for  kaons
and antikaons observed in heavy-ion collisions}

\author{
A.~F\"orster$^b$, F.~Uhlig$^b$, I.~B\"ottcher$^d$,
M.~D\c{e}bowski$^{e,f}$, F.~Dohrmann$^f$, E.~Grosse$^{f,g}$,
P.~Koczo\'n$^a$, B.~Kohlmeyer$^d$, F.~Laue$^{a,*}$, M.~Menzel$^d$,
L.~Naumann$^f$, H.~Oeschler$^b$, W.~Scheinast$^f$, E.~Schwab$^a$,
P.~Senger$^a$, Y.~Shin$^c$, H.~Str\"obele$^c$, C.~Sturm$^{b,a}$,
G.~Sur\'owka$^{a,e}$,
A.~Wagner$^f$, W.~Walu\'s$^e$\\
KaoS Collaboration\\
$^a$ Gesellschaft f\"ur Schwerionenforschung, D-64220 Darmstadt, Germany\\
$^b$ Technische Universit\"at Darmstadt, D-64289 Darmstadt, Germany\\
$^c$ Johann Wolfgang Goethe-Universit\"at, D-60325 Frankfurt am Main, Germany\\
$^d$ Phillips Universit\"at, D-35037  Marburg, Germany\\
$^e$ Jagiellonian University, PL-30059 Krak\'ow, Poland\\
$^f$ Forschungszentrum Rossendorf, D-01314 Dresden, Germany \\
$^g$ Technische Universit\"at Dresden, D-01062 Dresden, Germany\\
$^*$ Present address: Brookhaven National Laboratory,  USA }

\date{\today}

\begin{abstract}
Differential production cross sections of $K^-$ and $K^+$ mesons
have been measured in Ni+Ni and Au+Au collisions at a beam energy
of 1.5 $A\cdot$GeV. The $K^-/K^+$ ratio is found to be nearly
constant as a function of the collision centrality and system
size. The spectral slopes and the polar emission pattern differ
for $K^-$ and $K^+$ mesons. These observations indicate that $K^+$
mesons decouple earlier from the fireball than $K^-$ mesons.

\end{abstract}

\maketitle
%\begin{multicols}{2}
%\narrowtext

Heavy-ion collisions provide the unique possibility to study
baryonic matter well above saturation density. The conditions
inside the dense reaction zone and the in-medium properties of
hadrons can be explored by measuring the particles created in such
collisions \cite{aich,brown}. In particular, strange mesons
produced at beam energies below or close to the $NN$ threshold are
well suited for these studies. The yield of $K^+$ mesons as
measured in Au+Au collisions at SIS/GSI \cite{sturm} constrains
the nuclear matter equation-of-state \cite{fuchs}. The pronounced
patterns of the elliptic and directed flow of kaons provide
evidence for the existence of a repulsive kaon-nucleon in-medium
potential \cite{shin,crochet}. The $K^-$/$K^+$ ratio is enhanced
in heavy-ion collisions as compared to proton-proton collisions
\cite{barth,laue}. In order to reproduce the measured yields,
transport model calculations have to take into account
density-dependent $KN$ potentials corresponding to an in-medium
modification of the $K$ meson mass \cite{cass_brat,li_brown}. On
the other hand, the measured ratios of strange particles can be
explained within statistical models without any in-medium
modification of the masses. These models reproduce the measured
ratios by choosing an appropriate pair of values for the
temperature and the baryon chemical potential assuming thereby a
simultaneous chemical freeze-out~\cite{cley00}.

Quantitative information on the production mechanisms and the
properties of strange mesons in dense baryonic matter can be
extracted from the phase-space distributions of $K^+$ and $K^-$
mesons observed in heavy-ion collisions. In central Au+Au
collisions at beam energies above 4 $A\cdot$GeV the spectral
slopes were found to be similar for $K^+$ and $K^-$ mesons
\cite{Ahle}. The rapidity density distributions of strange mesons
from central Au+Au collisions at 10.7 $A\cdot$GeV and Pb+Pb
collisions at 158 $A\cdot$GeV were found to be wider for $K^+$
than for $K^-$ mesons \cite{AGS,NA49}. In both cases, the
$K^-/K^+$ ratio is constant as a function of the collision
centrality.  At beam energies below the $NN$ thresholds for
strangeness production ($NN \to K^+\Lambda N$ at E = 1.6 GeV, $NN
\to K^+ K^-NN$ at  E = 2.5 GeV), where in-medium effects are
expected to influence the kaon production significantly, a
systematic comparison of $K^+$ and $K^-$ phase-space distributions
has not yet been published.

In this Letter we present results of experiments on $K^+$ and
$K^-$ production in Ni+Ni and Au+Au collisions studied at a beam
energy of 1.5 $A\cdot$GeV. This is the lowest beam energy where
antikaons have been observed so far in collisions between heavy
nuclei. We have measured the spectral and angular distributions of
strange mesons as function of the collision centrality and have
found significant differences between kaons and antikaons.

The experiments were performed with the Kaon Spectrometer (KaoS)
at the heavy-ion synchrotron (SIS) at GSI in Darmstadt
\cite{senger}. Due to the energy loss in the Au target (thickness
0.5 mm) the average energy of the Au beam is 1.48 $A\cdot$GeV. The
energy loss of the Ni ions in the Ni target is negligible.  In
order to reach an energy of 1.5 $A\cdot$GeV for Au beams, an
exceptional operation of the GSI accelerator facility was
required: acceleration of the $^{197}$Au$^{63+}$ ions with the
synchrotron up to an energy of 0.3 $A\cdot$GeV, then extraction
and full stripping, then injection into the Experimental Storage
Ring (ESR) where the beam was cooled by electron cooling, then
re-injection into the synchrotron and acceleration up to 1.5
$A\cdot$GeV.

In order to study the centrality dependence we grouped the data
measured close to midrapidity ($\theta_{lab} = 40^{\circ}$) into
five centrality bins both for Ni+Ni and Au+Au collisions. The
centrality of the collision is derived from the multiplicity of
charged particles measured in the interval 12$^{\circ} <
\theta_{lab} < 48^{\circ}$. The most central collisions correspond
to 5\% of the total reaction cross-section $\sigma_R$, the
subsequent centrality bins correspond to 15\%, 15\% and 25\% of
$\sigma_R$. The most peripheral collisions correspond to 40\% of
$\sigma_R$. The total reaction cross-section has been derived from
a measurement with a minimum bias trigger and was found to be
$\sigma_R$ = 6.0$\pm$0.5 barn for Au+Au and $\sigma_R$ =
2.9$\pm$0.3 barn for Ni+Ni collisions. The corresponding number of
participating nucleons $A_{part}$ has been calculated from the
measured reaction cross-section fractions using a geometrical
model assuming a sharp nuclear surface.

\begin{figure}[hpt]
\epsfig{file=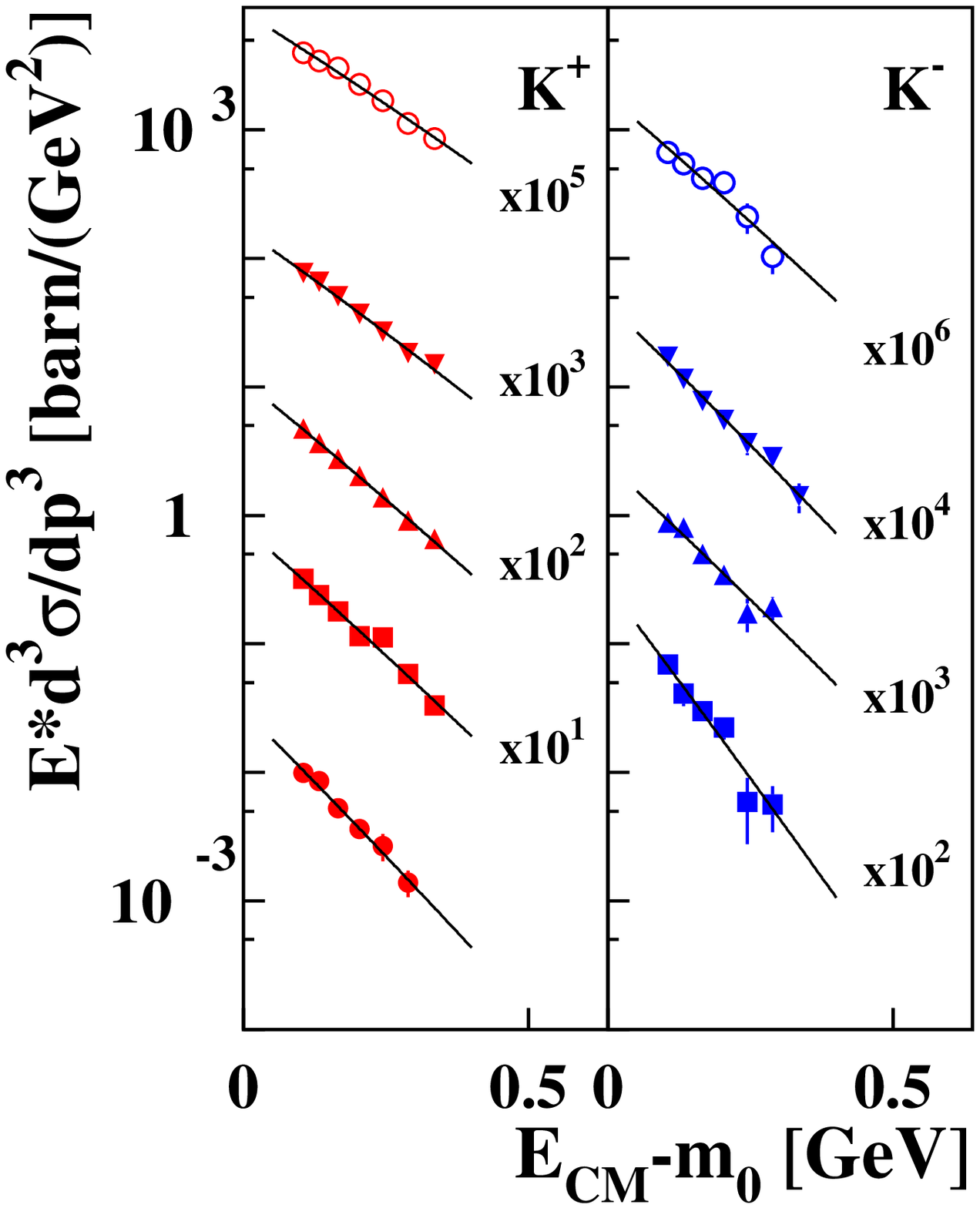,width=12.cm} \caption{Differential
production cross-sections for $K^+$ (left) and for $K^-$ mesons
(right) from Au+Au collisions at 1.5 $A\cdot$GeV for different
centrality bins as a function of the kinetic energy in the c.m.
system. The data were measured at an laboratory angle of
$\theta_{lab} = 40^{\circ}$ which covers midrapidity.  The spectra
correspond to bins of decreasing centrality (from top to bottom,
see text)} \label{spectra}
\end{figure}

Figure~\ref{spectra} shows the production cross sections for $K^+$
and $K^-$   mesons measured close to midrapidity as a function of
the kinetic energy in the center-of-momentum system  for the five
centrality bins in Au+Au collisions. The uppermost spectra
correspond to the most central reactions.  The error bars
represent the statistical uncertainties of the kaon and the
background events. An overall systematic error of 10\% due to
efficiency corrections and beam normalization has to be added. The
solid lines represent the function $ E \cdot d^3\sigma/dp^3 = C
\cdot E \cdot exp(-E/T)$ fitted to the data. $C$ is a
normalization constant and the exponential describes the energy
distribution with $T$ as the inverse slope parameter.

\begin{figure}
\epsfig{file=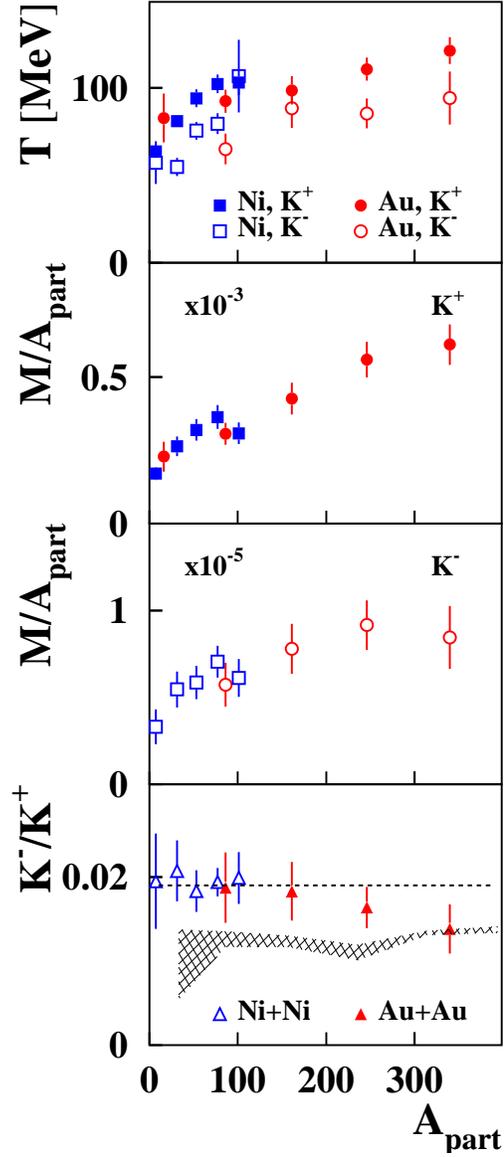,width=8.cm} \caption{First panel: Inverse
spectral slope parameters  of $K^+$ ( full symbols) and of K$^-$
mesons (open symbols) produced in Ni+Ni (squares)and Au+Au
 collisions (circles) at 1.5 $A\cdot$GeV as a function of the
number of participating nucleons. Second and third panel:
Multiplicity per number of participating nucleons $A_{part}$ of
$K^+$ and of $K^-$ as a function of $A_{part}$ both for Ni+Ni
(squares) and for Au+Au (circles) at a beam energy of 1.5
$A\cdot$GeV. The data were taken at a laboratory angle of
$\theta_{lab} = 40^{\circ}$. Lowest panel: The ratio of the $K^-$
to $K^+$ multiplicities as a function of $A_{part}$. The dashed
line represents the result of a statistical model calculation
\cite{cley00}. The cross-hatched area corresponds to the results
of a QMD transport model calculation for in-medium masses of $K$
mesons \cite{Hart02}. } \label{ratio}
\end{figure}

The spectra presented in Fig.~\ref{spectra} exhibit a distinct
difference between $K^-$ and $K^+$: The slopes of the $K^-$
spectra are steeper than those of the $K^+$ spectra. The inverse
slope parameters $T$ are displayed in the upper panel of
Figure~\ref{ratio} as a function of the number of participating
nucleons $A_{part}$. $T$ increases with increasing centrality and
is found to be significantly lower for antikaons than for kaons,
even for the most central collisions. When interpreting spectral
slopes one should keep in mind that they are influenced by both
the random and the collective motion of the particles (temperature
and flow). The radial-flow contribution to the slope depends on
the particle mass and hence cannot cause a difference between the
$K^+$ and the $K^-$ spectra. The temperature contribution to the
slope is determined at kinetic freeze-out, i.e. at the time when
the particles cease to interact.

The kaon multiplicity is defined for each centrality bin as $M =
\sigma_K/\sigma_r$ with $\sigma_K$ the kaon production cross
section and $\sigma_r$ the reaction cross-section of the
particular event class. Figure~\ref{ratio} presents $M/A_{part}$
for $K^+$ (second panel) and for $K^-$ (third panel) as a function
of $A_{part}$. Both for $K^+$ and $K^-$ mesons the multiplicities
exhibit a similar rise with $A_{part}$. Moreover, $M/A_{part}$ is
found to be almost identical in Ni+Ni and Au+Au collisions. The
$K^-/K^+$ ratio is about 0.02 below $A_{part} = 100$ and decreases
slightly to about 0.015 for the most central collisions
(Figure~\ref{ratio}, lowest panel).

\begin{figure}
\epsfig{file=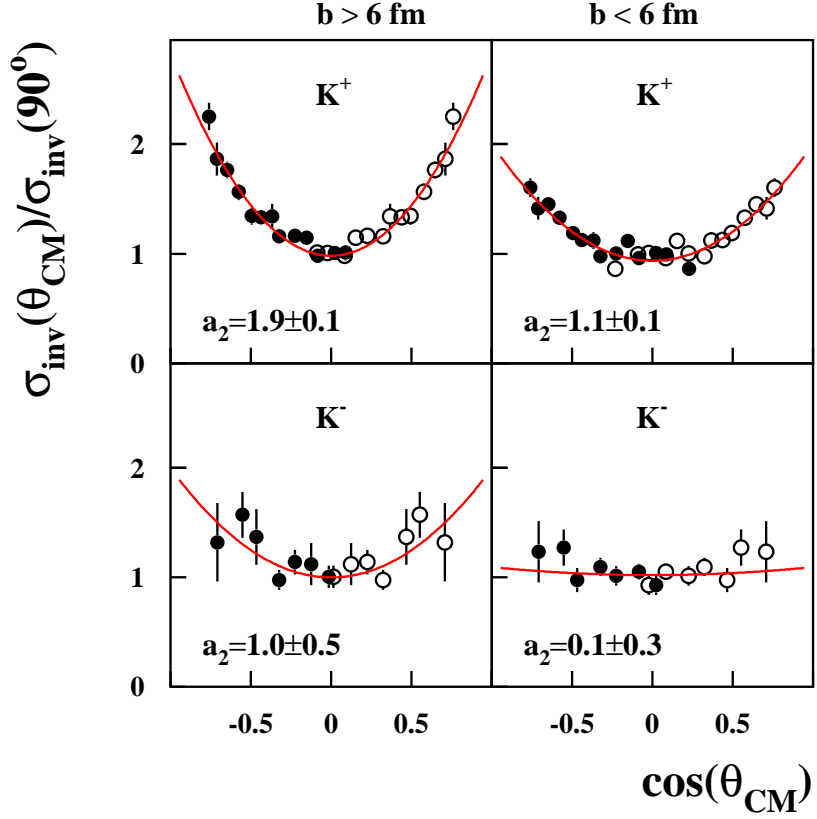,width=14.cm} \caption{Polar angle
distributions of $K^+$ (upper panel) and of $K^-$ mesons (lower
panel) produced in peripheral (left) and near-central (right)
Au+Au collisions at 1.5 $A\cdot$GeV. The data were taken at
laboratory angles between $\theta_{lab} = 32^{\circ}$ and
$72^{\circ}$ . Full data points are measured and mirrored at
$\cos(\theta_{CM})$ = 0 (open points). The lines correspond to the
function $1 + a_2 \cdot \cos^2(\theta_{CM})$ fitted to the data
(see text). The resulting values for $a_2$ are indicated.}
\label{polar}
\end{figure}

Another observable sensitive to the production mechanism is the
polar angle emission pattern.  The deviation from isotropy of the
$K^+$ and the $K^-$ emission can be studied by the ratio
$\sigma_{inv}(E_{CM},\theta_{CM})$/$\sigma_{inv}(E_{CM},
90^\circ)$ as a function of $cos(\theta_{CM})$. Here,
$\sigma_{inv}(E_{CM},\theta_{CM})$ is the invariant kaon
production cross-section measured at the polar angle $\theta_{CM}$
in the center-of-momentum frame and $\sigma_{inv}(E_{CM},
90^\circ)$ is the one measured at $\theta_{CM} = 90^\circ$. Due to
limited statistics we considered only Au+Au collisions grouped
into two centrality bins: near-central (impact parameter $b<$6 fm)
and non-central collisions ($b>$6 fm). Figure~\ref{polar} displays
the anisotropy ratio for $K^+$ (upper panel) and $K^-$ (lower
panel) and for near-central (right) and non-central collisions
(left). For an isotropic distribution this ratio would be constant
and identical to 1.

The solid lines in Fig.~\ref{polar} represent the function $1 +
a_2 \cdot \cos^2(\theta_{CM})$ which is fitted to the experimental
distributions with the values of $a_2$ given in the figure. In
near-central collisions the $K^-$ mesons  exhibit an isotropic
emission pattern whereas the emission of $K^+$ mesons is
forward-backward peaked. The angular distributions observed for
$K^+$ and $K^-$ in Ni+Ni collisions at 1.93 $A\cdot$GeV are
similar to the ones presented in  Fig.~\ref{polar} \cite{menzel}.
The measured emission patterns indicate that the antikaons - in
contrast to the kaons - have lost the memory of the beam direction
for central heavy-ion collisions.

In the following  we compare our data to the results of
theoretical calculations. Statistical models using a canonical
formulation of strangeness conservation predict a constant
$K^-/K^+$ ratio as a function of system size  for heavy-ion
collisions at SIS beam energies \cite{cley00}. The result of such
a calculation is shown in the lowest panel of Figure~\ref{ratio}
as a dashed line \cite{cley00}. In this case a baryochemical
potential of $\mu$ = 770 MeV and a chemical freeze-out temperature
of T = 63 MeV was assumed. Measured inverse slope parameters refer
to thermal freeze-out and are substantially larger: T($K^+$) =
103$\pm$6 MeV and T($K^-$) = 93$\pm$6 MeV for near-central Ni+Ni
collisions, and T($K^+$) = 116$\pm$7 MeV and T($K^-$) = 90$\pm$8
MeV  for near-central Au+Au collisions (corresponding to the
average value of the two most central bins in figure 2, upper
panel).

The observation of different spectral slopes or mean energy  for
$K^+$ and $K^-$ mesons  is at variance with a scenario in which
both kaons and antikaons have the same flow velocity and thermal
energy at chemical freeze-out. In consequence, the statistical
model does not offer a consistent explanation for both the yields
and spectral slopes of $K^+$ and $K^-$ mesons. The difference in
spectral slopes rather indicates that $K^+$ and $K^-$  mesons
decouple from the fireball sequentially due to their very
different KN inelastic cross sections.

Microscopic transport models predict that the kaons and the
hyperons are produced via processes like $NN\to K^+ YN$ or $\pi
N\to K^+ Y$ with $Y=\Lambda,\Sigma$ in the early phase of a
heavy-ion collision \cite{fuchs,aichelin,cass_brat}. The $K^+$
mesons leave the reaction volume with little rescattering because
of their long mean free path. Therefore, the $K^+$ mesons  probe
the early, dense and hot phase of the collision and have been used
to obtain information on the nuclear equation-of-state
\cite{sturm,fuchs}.  Within the transport calculations the
production of antikaons proceeds predominantly  via
strangeness-exchange reactions $\pi Y \to K^-N$
\cite{Ko,cass_brat,Hart02}. The mean free path of the $K^-$ mesons
is about 1.5 fm in nuclear matter due to absorption via reactions
like $K^-N\to Y\pi$. However, via the inverse reaction ($\pi Y \to
K^- N$) the antikaons may reappear again thus propagating to the
surface of the fireball. Consequently, the yields of $K^+$ and
$K^-$ mesons are both related to the hyperon yield, but the
observed $K^-$ mesons in average are produced later than the $K^+$
mesons \cite{Hart02}.

\begin{figure}
\epsfig{file=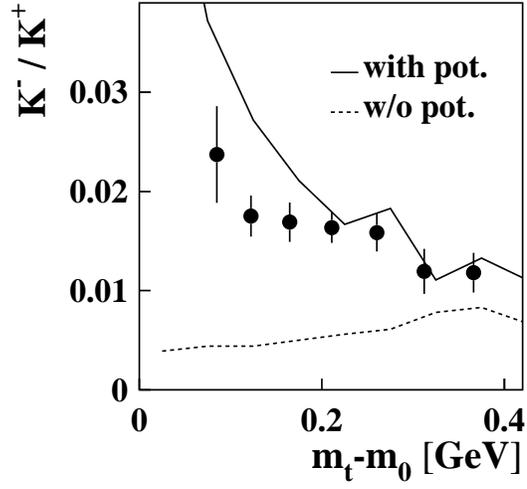,width=8.cm} \caption{$K^-$ to $K^+$ ratio
as a function of the transverse mass for near-central (b $<$ 5 fm)
Au+Au collisions at 1.5 $A\cdot$GeV around midrapidity. The curves
represent the predictions of a transport model calculation (BUU)
including in-medium potentials (solid line) and for free K meson
masses (dashed line). The data (full dots) were taken at
$\theta_{lab} = 40^{\circ}$ } \label{BUU}
\end{figure}

The various transport calculations do not yet provide a consistent
picture concerning the in-medium properties of antikaons. Recent
QMD model calculations predict a rather weak sensitivity of the
$K^-$ yield on the $K^-N$ potential \cite{Hart02}. The
calculations result in a $K^-/K^+$ ratio which systematically
underestimates the experimental data (see the hatched area in the
lowest panel of Figure~\ref{ratio})\cite{Hart02}. This result is
based on the assumption of in-medium $K^+$ and $K^-$ masses. A
very similar result is obtained for free masses. On the other
hand, BUU calculations need to take into account an attractive
in-medium $K^-N$ potential in order to explain the $K^-$ yields
\cite{cass_brat,li_brown}. Predictions of a BUU model calculation
\cite{li_brown} for the $K^-/K^+$ ratio as a function of
transverse mass for near-central Au+Au collisions at 1.5
$A\cdot$GeV are shown in Figure~\ref{BUU} together with our
experimental results. A similar result was found for Ni+Ni
collisions at 1.93 $A\cdot$GeV \cite{wisniewski}. As demonstrated
in Figure~\ref{BUU} the calculations assuming free K meson masses
(dashed line) and in-medium masses (solid line) clearly disagree.
In this case the differences in spectral slope are caused by the
opposite mean-field potentials of kaons and antikaons (see also
\cite{brat}). However, both the QMD \cite{Hart02} and the BUU
models \cite{li_brown,brat} use a rather simple parametrization of
the effective mass of $K^+$ and $K^-$ mesons in nuclear matter.
New theoretical concepts are required to improve the
interpretation of experimental data. This is expected from the
next generation of transport calculations which take into account
off-shell effects like in-medium spectral functions and in-medium
cross sections \cite{cass,lutz}.

In summary, we have presented differential cross sections and
phase-space distributions of  kaons and antikaons  produced in
heavy-ion collisions at 1.5 $A\cdot$GeV. We observed the following
features: (i) The $K^-/K^+$ yield ratio is quite independent of
$A_{part}$ both for Ni+Ni and Au+Au collisions,  (ii) in
near-central collisions $K^-$ mesons are emitted almost
isotropically whereas $K^+$ mesons exhibit a forward-backward
enhanced emission pattern, and (iii) the inverse slope parameters
are significantly smaller for $K^-$ than for $K^+$ mesons even for
the most central Au+Au collisions. These findings indicate that
(i) the production mechanisms of $K^+$ and $K^-$ mesons are
correlated by strangeness-exchange reactions, (ii)  $K^-$ mesons
undergo many collisions before leaving the fireball and, as a
consequence, (iii) $K^-$ and $K^+$ mesons experience different
freeze-out conditions.

We thank the GSI accelerator crew for an exceptional operation of
the GSI accelerator facilities resulting in a high-energy gold
beam of excellent quality. This work was supported by the German
Federal Government (BMBF), by the Polish Committee of Scientific
Research (No. 2P3B11515) and by the GSI fund for Universities.

\newpage

\end{document}